\pdfoutput=1
\documentclass[12pt,journal,comsoc]{IEEEtran}

\usepackage{soul}
\usepackage{color}
\usepackage{subfigure}
\usepackage{lipsum,graphicx,multicol}
\usepackage{multicol} 
\usepackage{caption}
\renewenvironment{IEEEbiography}[1]
  {\IEEEbiographynophoto{#1}}
  {\endIEEEbiographynophoto}
\usepackage{fancyhdr}

\usepackage{cite}

\ifCLASSINFOpdf
\else
\fi

\begin{document}
\chead{This work has been submitted to the IEEE for possible publication. Copyright may be transferred without notice, after which this version may no longer be accessible.
}

\title{BEAT: Blockchain-Enabled Accountable and Transparent Network Sharing in 6G}

\author{\IEEEauthorblockN{Tooba Faisal,
Mischa Dohler,
Simone Mangiante and 
Diego R. Lopez}}

\maketitle

\thispagestyle{fancy}

\begin{abstract}

Infrastructure sharing is a widely discussed and implemented approach and is successfully adopted in telecommunications networks today. In practice, it is implemented through prior negotiated Service Level Agreements (SLAs) between the parties involved. However, it is recognised that these agreements are difficult to negotiate, monitor and enforce. For future 6G networks, resource and infrastructure sharing is expected to play an even greater role. It will be a crucial technique for reducing overall infrastructure costs and increasing operational efficiencies for operators. More efficient SLA mechanisms are thus crucial to the success of future networks. 

In this work, we present ``BEAT'',  an automated, transparent and accountable end-to-end architecture for network sharing based on blockchain and smart contracts. This work focuses on a particular type of blockchain, Permissioned Distributed Ledger (PDL), due to its permissioned nature allowing for industry-compliant SLAs with stringent governance.  Our architecture can be implemented with minimal hardware changes and with minimal overheads.

\end{abstract}

\section{Introduction}

The deployment of 5G is well under way, but it is proving to be fairly challenging~\cite{gsma_inf_sharing}. Notably, with the yearly costs for 5G on operators expected to reach \$87.9 billion by 2023~\cite{heavy_reading}, challenges regarding site availability, acquisition and installation of the state-of-the-art equipment are proving to be a hindrance in the wide adoption of 5G services. This calls for radical changes in the way that current network infrastructure is managed from a deployment and operational point of view. With the main 3GPP 5G standard frozen with Release 16, there are opportunities to shape R17, R18 and subsequent 6G standards.


The experience of the 5G infrastructure rollout by the Mobile Service Providers (MSPs) provides a valuable base from which to start. 
Firstly, we note that 5G  operates at very high-frequency bands (i.e., 24 - 100 GHz), which allows operators to offer low latency services. The downside is that more cells are needed in order to continue to provide the same areal coverage as LTE. This translates into additional costs for the deployment of new cells, site acquisition and related infrastructure such as transmission lines. Given this context, the pooling of resources and network sharing between MSPs is an obvious way to reduce both CAPEX and OPEX. 


Indeed, and secondly, 5G inherently enables a multi-vendor environment such as OpenRAN, in which several vendors provide equipment for a single infrastructure. This is quite useful for several reasons, and particularly so with regards to infrastructure scaling without reliance on a single vendor. However, a multi-vendor environment imposes accountability and transparency requirements on equipment usage. From an accountability point of view, if a device malfunctions, network operators should be able to identify precisely the device which failed to meet its quality standards. Transparency, on the other hand, is essential because some dishonest vendors may  try to hide their errors and blame others, to save themselves from having to pay a penalty~\cite{vairam2019towards}.

In the 6G and beyond networks, the key for resource sharing is to enable trust among the network actors such as network operators and device vendors. Network sharing mechanisms proposed by research community tend to majorly focus on efficiency of resource sharing issues, for example \cite{crippa2017resource} discusses the efficient resource sharing in network slices and not accountable resource reservation at the device level. Samdani et el.\cite{samdanis2016network} rely on centralised entity for network sharing and neither provide any mechanism to maintain the record for future audit nor provides transparency to the network users for service provided.

The lack the ability to provide transparency and accountability, which we believe to be a significant oversight. A natively transparent, monitorable and accountable resource and infrastructure sharing architecture is essential for next generation systems.

To this end, this paper introduces our proposed solution to aforementioned problems and outlines a viable way forward for the industry. We introduce a blockchain-based end-to-end architecture for accountable and transparent (BEAT) sharing of network resources through smart contracts residing on Permissioned Distributed Ledgers (PDLs). PDLs are a particular type of distributed ledger that work between a closed group of mutually non-trusted parties. Access is managed via stringent access control mechanisms; i.e.~only authorised members can access a PDL, making them ideal for business-like applications. Furthermore, PDLs are immutable and contain executable smart contract code which gets deployed and executed on the ledger. 

The rest of the paper is organized as follows. A review of existing and related work is conducted in Section~\ref{sec:related_work}. In Section~\ref{sec:sys_arch2}, we present our first contribution "BEAT", a blockchain-based end-to-end architecture.~Our Second contribution is to evaluate our work in Section~\ref{sec:evaluation} by addressing the questions of the resource and performance overhead of PDLs. The paper ends with the concluding remarks of~Section~\ref{sec:conclusion}.

\section{Related Work}
\label{sec:related_work}

The use of blockchain-based distributed ledger technologies for network resource sharing is an emerging area of research that has been considered by several works recently. We restrict our discussion here to those results that are most closely related to our own efforts, and in particular, focus on resource sharing with distributed ledger and smart contract technologies.

A transparent on-the-fly Software Defined Network~(SDN) based technique for radio resource sharing architecture is proposed by~\cite{jiang2017radio}. In this work, a customer can connect to any available operator where the resources are available and is dependent on a third-party entity~(essentially an SDN-Server), which keeps track of available resources throughout the participating service providers. However, this work is limited to resource provisioning and neither discuss the prospects of low-level~(i.e., switch level) sharing nor accountability due to SLA violation.

Another Distributed Ledger Technology (DLT)-focused resource reservation work is \emph{Blockchain Network Slice Broker}~\cite{backman2017blockchain} and based on \emph{Network Slice Broker}\cite{samdanis2016network}. In that work, tenants (such as Over-the-Top providers) can request network services from the Mobile Network Operators~(MNOs) on-the-fly. The SLAs for the allocation is recorded  to a distributed ledger through smart contracts. However, the actual resource usage at a device is not recorded, and the problem of accountability in network sharing is not addressed.~Extension of Network Slice Broker~\cite{samdanis2016network} is presented in~\cite{NSB_chain} and provides an blockhain-focused architecture for network slice auction, in this work, infrastructure providers allocate network slices through an intermediate entity Intermediate Broker which further allocates resources to tenants. However, NSB doesn't discuss the problem of accountability.

Inter-operator network sharing architecture, specifically for smaller/denser cells, is proposed in~\cite{okon2020blockchain} and in \cite{mafakheri2018blockchain}. The network sharing SLAs between MNOs are stored as smart contracts on a distributed ledger and executed with service requests through an SDN layer.~A blockchain-focused unlicensed spectrum sharing approach is presented by~\cite{maksymyuk2019blockchain}. However this work discusses the prospect of network sharing among different players but doesn't provide any architecture of network sharing, however a game-theoretic algorithm for unlicensed spectrum sharing.
 

\section{Proposed System Architecture}
\label{sec:sys_arch2}

\begin{figure*}

  \includegraphics[width=\textwidth]{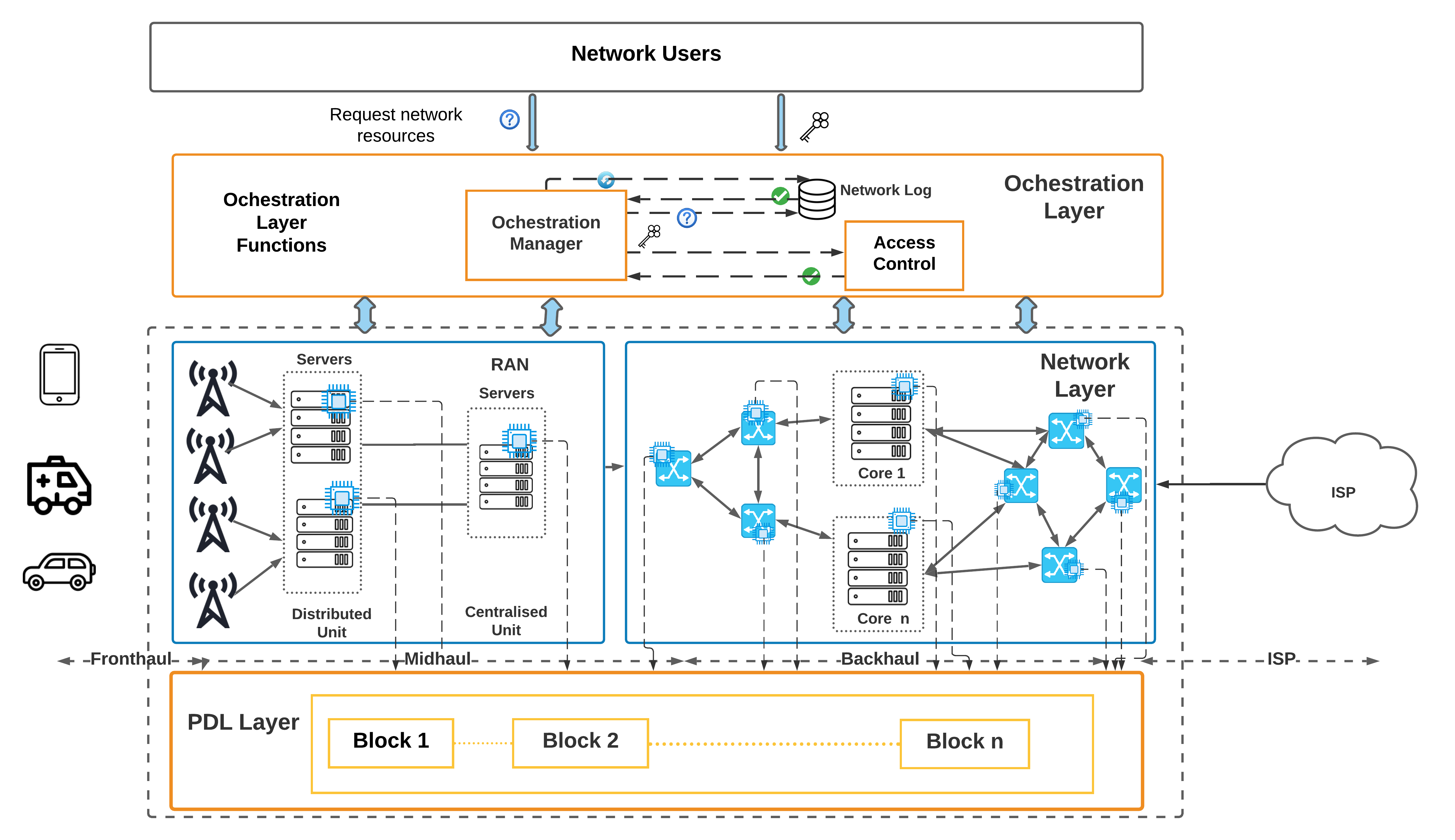}
  \caption{Network resource sharing architecture }
  \label{fig:system_arch_1}
\end{figure*}

BEAT is an automated architectural solution, which uses DLT and smart contracts to enables accountable and transparent multi-operator environment. It has three key elements: (1) Infrastructure sharing with (2) Distributed Ledgers and Smart contracts that enable (3) Accountability and Transparency. In this section, we explain the BEAT architecture in detail.


\subsection{BEAT's Architecture}

The architecture is underpinned by the following actors: \textit{\textbf{Network Users:}} They can be 1) \textit{Network Owners} -- A party or group of participants who own the infrastructure, or 2) \textit{Network Tenants} -- the party who lease the infrastructure from network owners, or 3) \textit{both }--  own some of the network infrastructure, which other tenants can lease. Moreover, they also  lease/rent some network infrastructure from the owners to serve their customers. \textit{\textbf{Device Vendors:}} provides network devices(e.g. routers and switches). 
\textit{\textbf{Governance:}}
The network governance is the consortium of the network users and include their representatives. It is up to the network users to decide the strategy (e.g., through voting) by which governance representatives are chosen.

Illustrated in Figure~\ref{fig:system_arch_1}, BEAT's architecture has three main operational layers: 1) an Orchestration Layer, 2) Network Layer, and 3) a PDL Layer. These are now discussed in subsequent sections.

\subsection{Orchestration Layer}

The \textbf{Orchestration Layer} is the top layer and handles the network resource requests from the tenants. Its operations are similar to ETSI's Management And Orchestration Layer~(MANO). This layer is maintained and managed by the governance of the PDL. It oversees the network operations and allocation decisions such as setting up network access, lease duration, price and privileges. Shown in Figure~\ref{fig:resource_alloc}, the Orchestration Layer has three main components:

\begin{figure}
    \centering
    \includegraphics[width=0.49\textwidth]{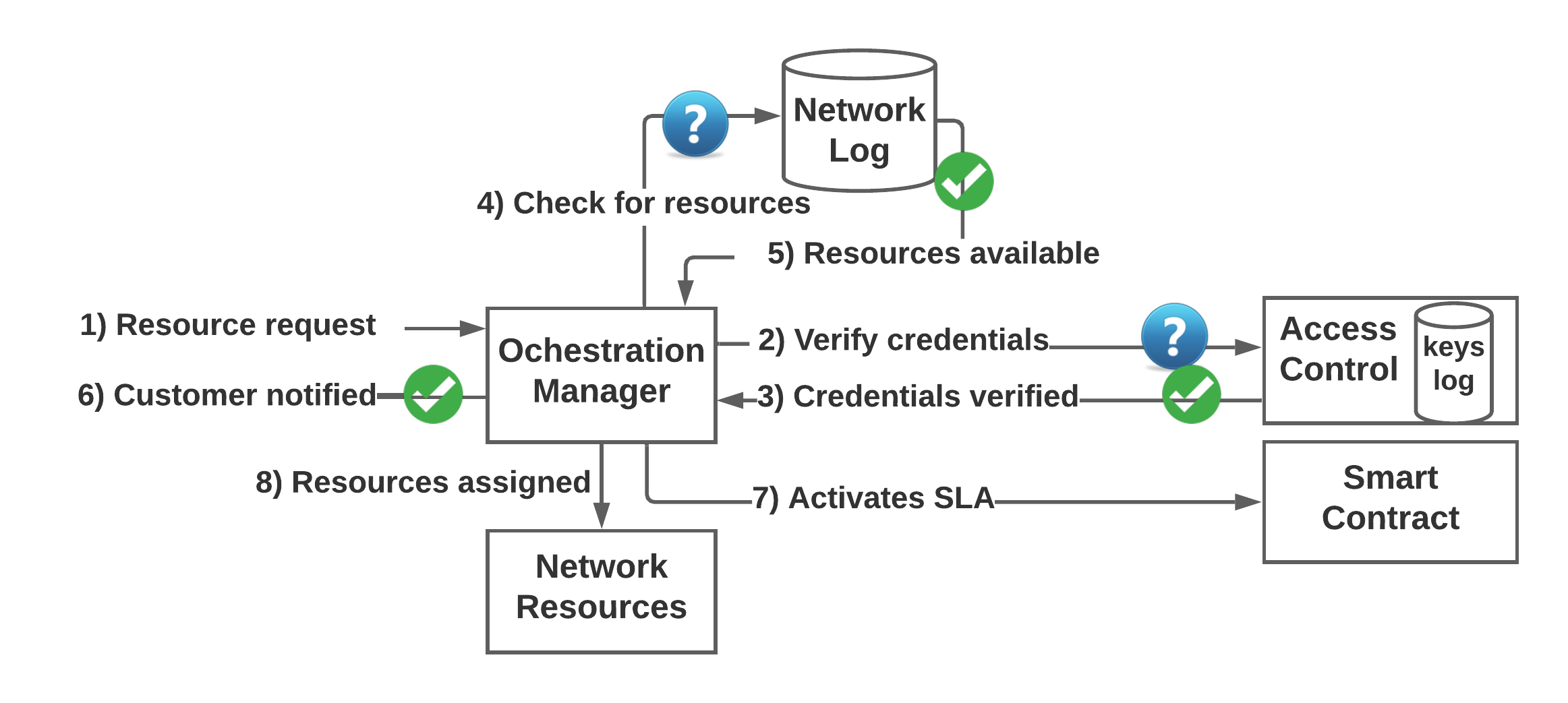}
    \caption{BEAT Orchestration Process}
    \label{fig:resource_alloc}
\end{figure}

 \emph{1)  Orchestration Manager:} It serves the incoming requests and has  features such as \emph{Universal View} similar to the Software-Defined Mobile Network Orchestrator~(SDM-O) discussed in~\cite{crippa2017resource} and the  SDN-Server of~\cite{jiang2017radio}. When a network participant joins the network, the Orchestration Manager assigns the credentials and keeps the record. The Orchestration Manager also allocates a node PDL-ID to a device. This PDL-ID is different from the Layer 2/3 addresses because the devices may change their IP addresses anytime and the PDL-ID must remain same.
 
\emph{2) Access Control:} An access control verification entity which maintains a database to keep the record of credentials and replies to  access control confirmation queries from the Orchestration Manager.

\emph{3) Network Log:}  A database to maintain network resource logs.

\subsection{Network and PDL Layers}
Infrastructure and network resources, such as switches and routers, form the \textbf{Network Layer} of BEAT; it is managed and maintained by its Orchestration Layer. 


When the tenants request network resources, the Orchestration Manager will verify from the access control entity if the tenant has an agreement already. If the agreement is present, the Orchestration Manager will then check from the \emph{Network Log} the status of the network, that is, the current load on each path of the network and on each device. If the network has resources available, the Orchestration Manager will send a confirmation message to the tenant. Next, a smart contract will be executed to initialize the SLA and then orchestrate the network resource for the tenant.~If the capacity is not available, the tenant can wait for the resources to become available.

BEAT advocates a multi-operator and multi-vendor environment. Therefore, stakeholders must know the performance and usage of the network components at a very fine-grained level. Such performance metrics are required for future SLA compliance and accountability of the sharing agreement. The infrastructure usage in BEAT is recorded at device level and transparently shared with smart contracts at the \textbf{PDL Layer}. Every device in BEAT is equipped with a PDL node and can execute smart contracts to record relevant data to the PDL. At macro-level, all the devices together form a PDL within the network infrastructure.

\subsection{Automated Recording with Smart Contracts}
A tenant's main objective is to get agreed on service levels to serve its customers' demands. Service quality can get affected for several reasons, such as a slower path or network device malfunctioning. In the situations of degraded service, the tenant should get compensation if applicable and without any hassle. In such cases, all the stakeholders (i.e., network operators and vendors) would blame each other to avoid paying the penalty. Therefore, there should be a mechanism to record the service data, which no party can deny.

In BEAT, the flows' data is recorded to the PDL through smart contracts. For each flow, the source and the destination both records the relevant data to the PDL. PDLs are immutable, which means data recorded to them cannot be deleted. Moreover, PDLs are transparent, and all the participants for the consortium can see the flow source and destination information in the PDL. To resolve these two problems of scalability and privacy, we install minimum hashed data per flow to the PDL; specifically, \emph{node ID}, \emph{source IP address}, \emph{destination IP address} and \emph{timestamp}. A smart contract is executed by the source and destination switches and record the hashed data~(i.e., $SHA3(node\_ID, src\_IP , dst\_IP,time\_stamp)$) to the PDL.~Here the Source IP refers to Source of flow or first switch for the flow and destination IP is the final switch/device the packet is destined to.

\begin{figure}
    \centering
    \includegraphics[width=0.49\textwidth]{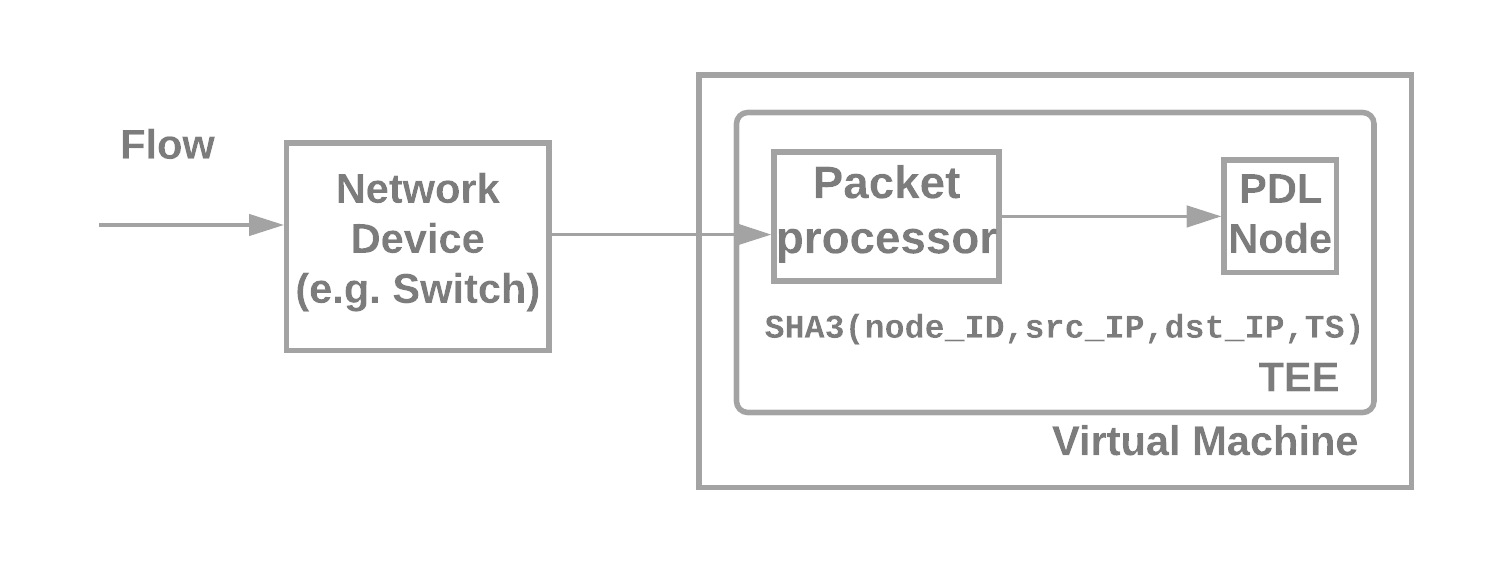}
    \caption{The Packet Processor -- records the per flow packet data (i.e., \emph{node\_ID, src\_IP, dst\_IP, timestamp}) to the PDL through a smart contract.}
    \label{fig:packet_processor}
\end{figure}

\subsection{TEEs for Flow Monitoring}
\label{subsec:TEE}

The data is recorded to the PDL by means of a smart contract that is executed by the \emph{Packet Processor} -- which is a software script that runs on a virtual machine within the PDL node adjacent to the device; see Figure~\ref{fig:packet_processor}.  The packet processor extracts the data from the flow, hashes it and executes a smart contract to record the data to the PDL node. Packets information recorded through smart contract enables accountability. In BEAT, the owner has no control over this virtual machine -- enabling secured and trustworthy data recording.

However, both the Packet Processor and the PDL node are installed on network owner controlled devices, and the tenants may not trust them as owners could tamper with data. 
Such as they may record a delayed packet receipt time or deny the receipt of the packet altogether. It is difficult to dispute such claims due to the best-effort nature of today’s internet. 

Therefore, BEAT adds another layer of security and wraps the Packet Processor and the PDL node inside a Trusted Execution Environment (TEE).  It is to be noted here that TEE is a separate secure processing system that solves this trust issue. Note that the packets can still be dropped in the network but recording the  packet receipt at the source device ensures that packet was sent correctly. The governance can give the tenants controlled access to the virtual machine -- this access is decided by the PDL voting mechanisms and will rotate among the tenants to enable trustworthy record keeping. 
\section{Evaluating BEAT}
\label{sec:evaluation}
\begin{figure*}
\subfigure[Simulation Setup in GNS3 ]{\label{fig(a):simulation_setup}\includegraphics[width=0.33\textwidth]{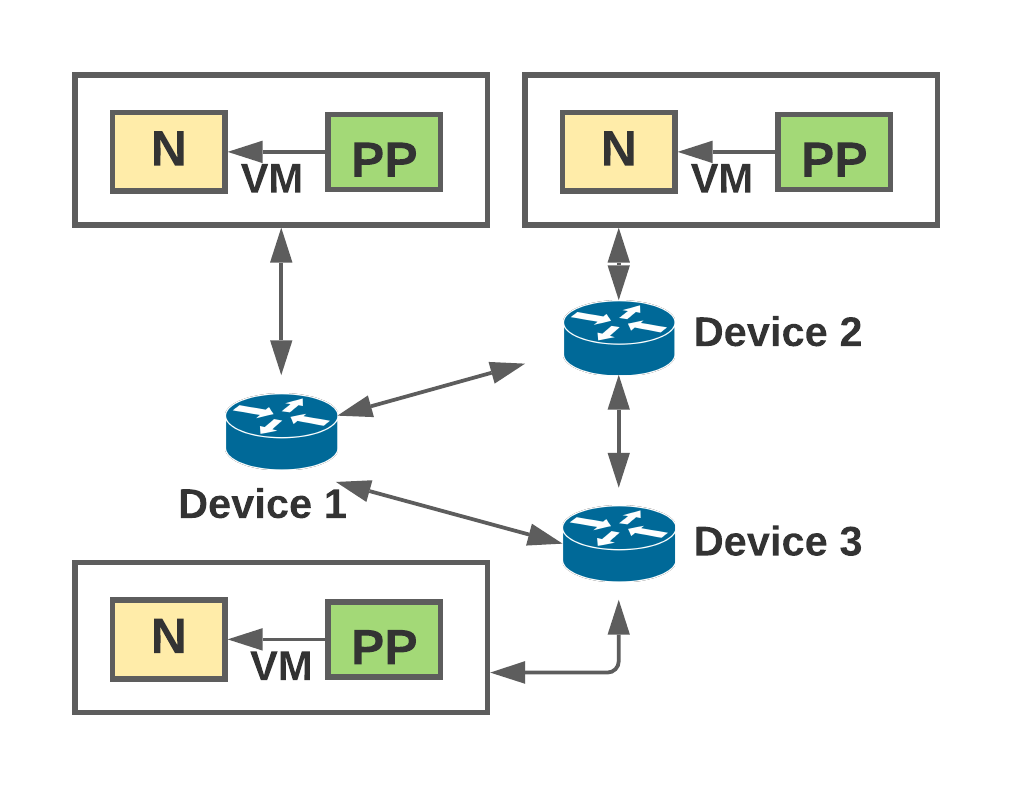}}\subfigure[Contract Execution Time]{\label{fig:(b)contract_inv}\includegraphics[width=0.33\textwidth]{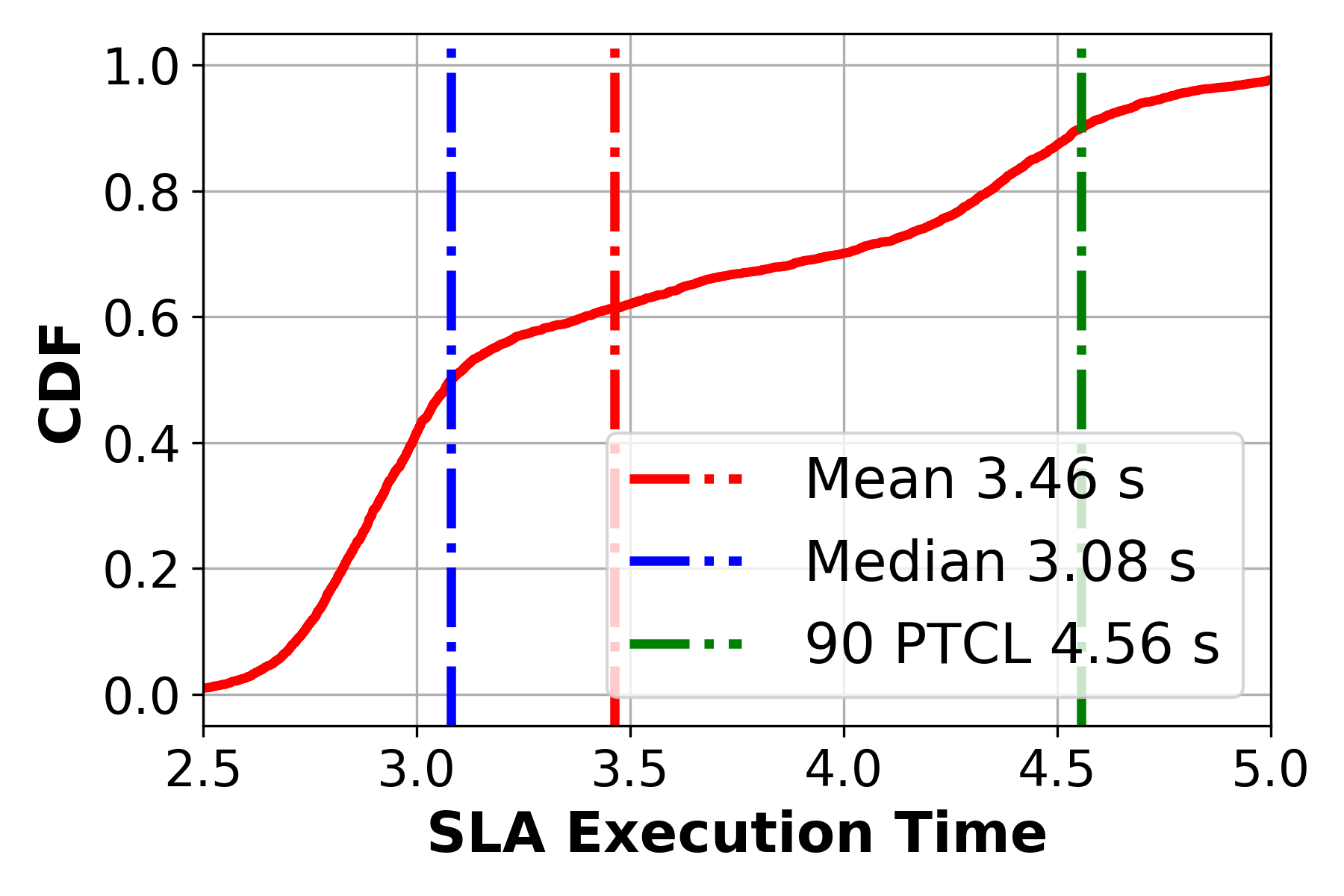}}\subfigure[CPU Utilisation]{\label{fig:(c)cpu_utilization}\includegraphics[width=0.33\textwidth]{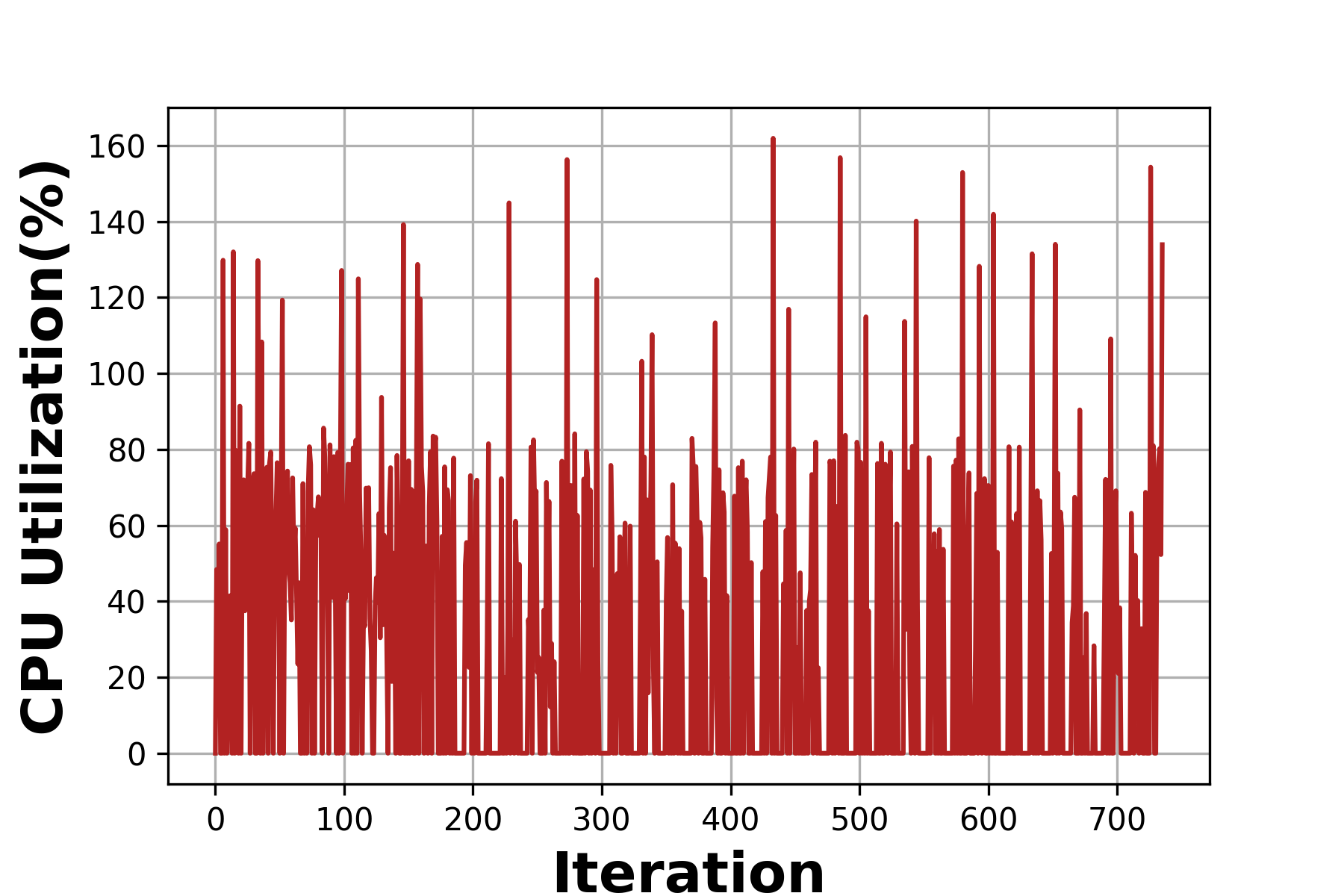}}

\caption{BEAT Evaluation -- a) shows the simulation setup with PDL Node (N) and Packet Processor (PP). b) shows the smart contract (SLA) execution time and c) shows the CPU overhead due to BEAT. Our results show that BEAT adds a negligible~(around 4 seconds) overhead altogether to data recording. Note that BEAT records the data to the ledger through a smart contract and \emph{only} at the start and at the end of the flow.}

\end{figure*}

The aim of our evaluation is to establish the overheads introduced by BEAT. To this end, we installed PDL nodes on every network device within the GNS3 simulated network infrastructure that formed an end-to-end PDL system.

We studied the compatibility between two independent systems, that is, the PDLs and the network infrastructure. We evaluated the viability of our proposal with a simple but similar real-world network topology~(Figure~\ref{fig(a):simulation_setup}).
The idea of this study was to enable accountability with PDLs at the network layer - therefore, we installed one Ethereum node on each of the three edge routers. We advocated the use of Permissioned Ledger -- hence we adopted Ethereum's permissioned version with Proof-of-Authority~(PoA) consensus protocol; more specifically, the ``Clique'' implementation of PoA which has higher throughput and lower latency than traditional Proof-of-Work protocol~\cite{singh2019managing} and blocks can be generated at a user-defined time interval. Clique also outperforms its other PoA counterparts and requires less message exchanges and hence delivers high throughput as compared to other permissioned flavours of Ethereum such as Aura~\cite{de2018pbft}.

We set up our simulations on an Intel Core i7 CPU 2.70 GHz with 16 GB RAM running Windows 10 OS; with GNS3 network simulator version 2.2.17. The router images in GNS3 are using Cisco IOSv 15.7(3)M-2. The \emph{packet processor} shown in Figure~\ref{fig:packet_processor} is coded with Python 3  and installed on a virtual machine with the PDL node. The blocks for the PDL are generated at 15 seconds intervals. We have used 15 seconds as a benchmark, and in the production environment, the block generation time will depend on several factors such as PDL type and the number of devices in the network. The smart contract which represents the SLA and to record the usage is coded in Solidity and installed on the ledger. The smart contract compiled in average approximately 0.11 seconds and deployed in approximately 14 seconds average. 

The Packet Processor processes the flow data, that is, \emph{node ID}, \emph{source IP address}, \emph{destination IP address} and \emph{timestamp}. to In our simulations, it took average 0.65 ms to capture the packet's data. The Packet Processor then, hashes this data and records it to the PDL through smart contract execution. We advocate the use of plugable hashing algorithm -- that is, the governance of the PDL has the liberty to choose their preferred algorithm based on their priorities, such as the availability of computational power.

In this work, however, we chose the 32-byte version of SHA3, that is, SHA3-256. In our experiments, it took $\approx$ 0.31 ms to hash the 46  to 56 bytes processed data.
The next step was to install this hashed data to the PDL; this was done through a pre-installed smart contract execution and took order of seconds to execute(approximately 4 seconds); see Figure~\ref{fig:(b)contract_inv}. 

The CPU utilization due to our scripts is approximately 30\% mean (Figure~\ref{fig:(c)cpu_utilization}). The memory usage is around 44 MB for the script in our simulations.  Note that we ran these experiments in a local laptop and a GNS3 simulator which runs another virtual machine to manage the network devices. The resources in our simulation test-beds are shared among the OS, GNS3 simulator, virtual machine and the containers managed by the simulator. With modern network switches~\cite{cisco_wp}, which can have high speed CPUs with up to 28 cores, we anticipate a far improved performance in production. 

With these experiments, we observe that BEAT's additional overhead is less than 4 seconds. This means that \emph{our automated, transparent and accountable sharing mechanism  results in negligibly small additional processing times  for the stakeholders and, most importantly, with minimal hardware changes.}

\section{Considerations}
\label{sec:sla_violation}

Designing a system with PDLs is not trivial. They have several considerations, such as transaction throughput and scalability.~In this section, we discuss the considerations related to the BEAT.

\paragraph{Integrity of Data}
Smart contracts do not have any built-in means to verify the integrity of the data. Hence, it is vital to ensure that the data recorded by network devices is valid. 
In BEAT, TEE ensures that the correct data is recorded to the PDL. However, the network device feeds the data to the packet processor and if this device is malicious, it will provide false information to the PDL.

This is one of the reasons we advocated the use of governance-controlled \emph{Permissioned} Distributed Ledgers. In a PDL, the members are allowed with access control mechanisms and affected parties can report such misbehaviour to the governance, which can take subsequently disciplinary actions and \emph{blacklist} the node and may impose penalties.

\paragraph{Colluding MSPs}

In a Blockchain-enabled architecture like BEAT, the MSPs can collude with each other. In such a case, dominant MSPs can behave maliciously towards other tenants such as reject their transactions. To solve such problems, regulatory authority~(e.g. Ofcom in the UK and Federal Communications Commission (FCC) in the US) can also be the part of the PDL network governance. Note that, the role of regulatory authority is an observation role only and should be contacted in the situations of dispute only. The regulatory authority neither take part in consensus nor control any device. 

\paragraph{Denial-Of-Service~(DoS)}
Every PDL allows a particular number of Transactions Per 
Second~(TPS), which are generally higher in PDLs (e.g., 20,000 TPS~\cite{gorenflo2020fastfabric}) than permissionless ledgers~(e.g., Ethereum $\approx$ 10 TPS). This means that if many devices send data simultaneously, it can cause congestion at the ledger or DoS for incoming transactions. Therefore, whilst designing a PDL network, it is important to adopt a PDL-type (e.g., Hyperledger Fabric), which can cope with the network's requirement.

\paragraph{Routers' Fraud}
Routers are vendor or network owner-controlled devices, and may send the users' data to a slower and cheaper path rather than a faster and agreed path. In BEAT, we proposed to record the data on the source and the destination device only to solve the scalability and congestion problems. Note that BEAT records the \emph{timestamp(ts)}  for each flow, both at the source and the destination. This means, the latency-focused SLA can be estimated with this information (e.g. \emph{$ts_{end}$ - $ts_{start}$}). 

However, it is also possible to record the complete path of the packet with BEAT and record the data on every device if scalability and congestion of the PDL is not a concern.
\paragraph{Waiting Times}
Recall that the Orchestration Manager allocates the resources considering the available capacity of the network. Therefore, it is likely that some users will have to wait to get hold of the resources. BEAT is an accountability focused architecture that is adherence to the SLA.~Controlled resource allocation ensures that the network is not overwhelmed by the service requests and users receive SLA promised services.

\section{Conclusion and Future Work}
\label{sec:conclusion}

In this work, we presented ``BEAT'', a PDL focused automated, transparent,  accountable network sharing architecture operating at the network layer. 
BEAT is a layered architecture in which governance of the PDL manages and maintains the network resources with stringent access control and network management strategies; to ensure SLA guarantee.

BEAT adds a negligible overhead  to the system. When considering the time and cost required for the negotiation of SLAs for infrastructure sharing. We believe our system provides a faster and more-seamlessly automated solution. 

For now, the smart contracts are pre-defined; in our future work, we intend to explore the possibility of \emph{Intelligent Smart Contracts}, which can adapt and change the parameters in line with changing traffic conditions. Also, in this work, we proposed BEAT functionality on a VM. However, in future, we envision network devices with built-in BEAT functions.

In conclusion,  it is our hope that this work  marks  the  inception of a new era of network sharing in which competing stakeholders can work efficiently and transparently.

\bibliographystyle{IEEEtran}
\bibliography{bib.bib}

\begin{thebibliography}{10}
\providecommand{\url}[1]{#1}
\csname url@samestyle\endcsname
\providecommand{\newblock}{\relax}
\providecommand{\bibinfo}[2]{#2}
\providecommand{\BIBentrySTDinterwordspacing}{\spaceskip=0pt\relax}
\providecommand{\BIBentryALTinterwordstretchfactor}{4}
\providecommand{\BIBentryALTinterwordspacing}{\spaceskip=\fontdimen2\font plus
\BIBentryALTinterwordstretchfactor\fontdimen3\font minus
  \fontdimen4\font\relax}
\providecommand{\BIBforeignlanguage}[2]{{%
\expandafter\ifx\csname l@#1\endcsname\relax
\typeout{** WARNING: IEEEtran.bst: No hyphenation pattern has been}%
\typeout{** loaded for the language `#1'. Using the pattern for}%
\typeout{** the default language instead.}%
\else
\language=\csname l@#1\endcsname
\fi
#2}}
\providecommand{\BIBdecl}{\relax}
\BIBdecl

\bibitem{gsma_inf_sharing}
GSMA, ``{Infrastructure Sharing: An Overview},'' \url{https://bit.ly/2QWGR6c},
  {Accessed on: 27 March 2021}.

\bibitem{heavy_reading}
{Heavy Reading}, ``{Mobile Operator 5G Capex Forecasts: 2018 - 2023},''
  \url{https://bit.ly/3cNE9IR}, {Accessed on: 27 March 2021}.

\bibitem{vairam2019towards}
P.~K. Vairam, G.~Mitra, V.~Manoharan, C.~Rebeiro, B.~Ramamurthy \emph{et~al.},
  ``{Towards Measuring Quality of Service in Untrusted Multi-Vendor Service
  Function Chains: Balancing Security and Resource consumption},'' in
  \emph{IEEE INFOCOM}, 2019, pp. 163--171.

\bibitem{crippa2017resource}
M.~R. Crippa, P.~Arnold, V.~Friderikos, B.~Gajic, C.~Guerrero, O.~Holland,
  I.~L. Pavon, V.~Sciancalepore, D.~von Hugo, S.~Wong \emph{et~al.},
  ``{Resource Sharing for a 5G Multi-tenant and Multi-service Architecture},''
  pp. 1--6, 2017.

\bibitem{samdanis2016network}
K.~Samdanis, X.~Costa-Perez, and V.~Sciancalepore, ``{From network sharing to
  multi-tenancy: The 5G network slice broker},'' \emph{{IEEE Comms. Mag.}},
  vol.~54, no.~7, pp. 32--39, 2016.

\bibitem{jiang2017radio}
M.~Jiang, D.~Xenakis, S.~Costanzo, N.~Passas, and T.~Mahmoodi, ``{Radio
  Resource Sharing as a Service in 5G: A Software-Defined Networking
  Approach},'' \emph{Computer Communications}, vol. 107, pp. 13--29, 2017.

\bibitem{backman2017blockchain}
J.~Backman, S.~Yrj{\"o}l{\"a}, K.~Valtanen, and O.~M{\"a}mmel{\"a},
  ``{Blockchain Network Slice Broker in 5G: Slice Leasing in Factory of the
  Future Use Case},'' pp. 1--8, 2017.

\bibitem{NSB_chain}
L.~Zanzi, A.~Albanese, V.~Sciancalepore, and X.~Costa-Pérez, ``{NSBchain: A
  Secure Blockchain Framework for Network Slicing Brokerage},'' in \emph{{IEEE
  ICC}}, 2020, pp. 1--7.

\bibitem{okon2020blockchain}
A.~Okon, N.~Jagannath, I.~Elgendi, J.~M. Elmirghani, A.~Jamalipour, and
  K.~Munasinghe, ``{Blockchain-Enabled Multi-Operator Small Cell Network for
  Beyond 5G Systems},'' \emph{IEEE Network}, vol.~34, no.~5, pp. 171--177,
  2020.

\bibitem{mafakheri2018blockchain}
B.~Mafakheri, T.~Subramanya, L.~Goratti, and R.~Riggio, ``{Blockchain-Based
  Infrastructure Sharing in 5G Small Cell Networks},'' in \emph{{IEEE CNSM}},
  2018, pp. 313--317.

\bibitem{maksymyuk2019blockchain}
T.~Maksymyuk, J.~Gazda, L.~Han, and M.~Jo, ``{Blockchain-based Intelligent
  Network Management for 5G and Beyond},'' in \emph{{IEEE AICT}}, 2019, pp.
  36--39.

\bibitem{singh2019managing}
P.~K. Singh, R.~Singh, S.~K. Nandi, and S.~Nandi, ``Managing smart home
  appliances with proof of authority and blockchain,'' in \emph{I4CS}, 2019,
  pp. 221--232.

\bibitem{de2018pbft}
S.~D. Angelis, L.~Aniello, R.~Baldoni, F.~Lombardi, A.~Margheri, and
  V.~Sassone, ``{PBFT vs Proof-of-Authority: Applying the CAP Theorem to
  Permissioned Blockchain},'' in \emph{ITASEC}, 2018.

\bibitem{cisco_wp}
``{Second-Generation Intel Xeon Scalable Processor Refresh Selection Guide for
  VDI on Cisco UCS with VMware Horizon 7},'' \url{https://bit.ly/3CQhENz},
  {Accessed on: 27-09-2021}.

\bibitem{gorenflo2020fastfabric}
C.~Gorenflo, S.~Lee, L.~Golab, and S.~Keshav, ``{FastFabric: Scaling
  hyperledger fabric to 20,000 transactions per second},'' \emph{Int. J. Netw.
  Manag.}, vol.~30, no.~5, p. e2099, 2020.

\end{thebibliography}
\vspace{-1pt}
\begin{IEEEbiography}{Tooba Faisal} is a PhD student at King's College London (KCL) working on Telco-blockchain. Her current research interests are designing secure smart contracts, Distributed Ledger Technology and service level agreements (SLAs). She also has a Master of Research (MRes) in Security Science from University College London, a Master of Science (MS) in Telecommunication and Networks and a Bachelor of Science degree in Computer Engineering from Bahria University, Karachi, Pakistan. She is a KCL's delegate in the ETSI Industries Specifications Group on Permissioned Distributed Ledgers and rapporteur of smart contract Group Report and Specifications. 
\end{IEEEbiography}
\vspace{-1cm}
\begin{IEEEbiography}{Mischa Dohler} is the Chief Architect in Ericsson and visiting Professor in Wireless Communications at King’s College London, driving cross-disciplinary research and innovation in technology, sciences, and arts. He is a Fellow of the IEEE, the Royal Academy of Engineering, the Royal Society of Arts (RSA), the Institution of Engineering and Technology (IET); and a Distinguished Member of Harvard Square Leaders Excellence. He sits on the Spectrum Advisory Board of Ofcom, and acts as policy advisor on issues related to digital, skills and education. 
\end{IEEEbiography}
\vspace{-1cm}
\begin{IEEEbiography}{Simone Magiante} received his Ph.D. in computer networks in 2013 from the University of Genoa, Italy, working on carrier Ethernet management using the SDN paradigm. He then spent three years with Dell EMC in Ireland as a senior research scientist, where he managed European projects focusing on SDN and network transport. He led the design and deployment of a virtualized industrial IoT testbed and contributed to several H2020 EU proposals. He is currently a research and standards specialist in Vodafone Group, United Kingdom. 
\end{IEEEbiography}
\vspace{-1cm}
\begin{IEEEbiography}{Diego R. Lopez}
 joined Telefonica I+D in 2011 as a Senior Technology Expert on network middleware and services. He is currently in charge of the Technology Exploration activities within the GCTO Unit of Telefónica I+D. Before joining Telefónica he spent some years in the academic sector, dedicated to research on network service abstractions and thedevelopment of APIs based on them. During this period he was appointed as member of the High Level Expert Group on Scientific Data Infrastructures by the European Commission.~Diego is currently focused on identifying and evaluating new opportunities in technologies applicable to network infrastructures, and the coordination of national and international collaboration activities. 
\end{IEEEbiography}

\end{document}